\documentclass[aps,prb,twocolumn,groupedaddress,showpacs,showkeys]{revtex4-1}
\usepackage{graphics}
\newcommand{\ud}{\mathrm{d}}
\begin{document}
\title{Correlations of superconducting fluctuations in a two-gap system}

\author{Artjom Vargunin}
\author{Teet \"Ord}
\email[]{teet.ord@ut.ee}
\affiliation{Institute of Physics, University of Tartu, 4 T\"ahe Street,
51010 Tartu, Estonia}

\date{\today}

\begin{abstract} We derive analytically the spatial correlation functions for gap fluctuations in two-band scenario with intra- and interband
pair-transfer interactions. These functions demonstrate the changes in spatial functionality due to the presence of two channels of coherency described by the divergent and finite correlation lengths. Even at the phase transition point both channels are essential for two-band superconductivity. Generally their relative contributions depend on the temperature and system parameters.
\end{abstract}


\maketitle


%
%

\section{Introduction} The theory of superconductivity with overlapping bands has started to develop since 1959 \cite{suhl, *moskalenko}, however, only after discovery of multicomponent nature of MgB$_2$ in 2001 \cite{MgB0a,*MgB0b,*MgB0c,*MgB0d} and pnictides in 2008 \cite{pnictides0a,*pnictides0b,*pnictides0c} the multigap approaches have become an object of exceeding interest.

The peculiarities of the spatial coherency in multiband superconductors have attracted much attention recently in connection with type-$1.5$ behaviour \cite{babaev0,*moshchalkov0}. In usual one-band systems there is only one coherence length which value in units of penetration depth determines response to magnetic fields, type-I or type-II. It was suggested that in two-band superconductor one has two correlation lengths resulting much richer physics than type-I/type-II dichotomy. In particular, there is a possibility to observe a mixture of domains of Meissner state and vortex clusters, called type-1.5 superconductivity. The latter regime is supported also by interband proximity effect \cite{babaev3,*babaev4} and by different kinds of intercomponent interaction involving Josephson, mixed gradient or density-density couplings \cite{babaev6}.

The existence of two qualitatively different length scales in two-band system was demonstrated more than twenty years ago \cite{poluektov} and recently \cite{babaev2,ord1}. Two distinct correlation lengths are also present in negative-$U$ Hubbard model of two-orbital superconductivity \cite{ord3,*ord4}. In this respect the connection between peculiarities of spatial coherency and excitation of the Leggett mode in two-gap material was discussed \cite{ord2}.

Different point of view on the correlation behaviour in two-band model is based on the statement that within the Ginzburg-Landau domain both order parameters vary on the same length scale \cite{kogan,kogan2}. An extension of the temperature domain indicates that two gaps are generally not proportional to each other, thus their spatial scales are decoupled only away from critical point \cite{shanenko2}. Numeric estimations for the healing lengths of the gaps confirm that conclusion for several superconducting materials \cite{shanenko3}. Here we note that scientific discussion about discrepancy of length scales in the vicinity of critical temperature still continues \cite{babaev5,*kogan3}.

Experimentally, two characteristic length scales in various two-band compounds were evidenced e.g. by vortex imaging \cite{exp1}, muon spin relaxation measurements \cite{exp2}, in heat transport features \cite{exp3} as a function of magnetic field.

In the present contribution we derive correlation functions for gap fluctuations in the two-band scenario. The spatial behaviour of these characteristics reveal two different correlation lengths describing joint superconducting condensate as a whole. These length scales are analysed as the functions of the temperature and interband interaction constant. The competition between the contributions of the corresponding coherency channels to the correlation functions are discussed.

\section{Derivation of correlation functions} We start with two-band superconductivity Hamiltonian
\begin{eqnarray}\label{e1}
&&H=\sum_{\alpha\mathbf{k}s}\tilde{\epsilon}_\alpha(\mathbf{k})a^+_{\alpha\mathbf{k}s}a_{\alpha\mathbf{k}s}\nonumber\\
&&-\frac{1}{V}\sum_{\alpha\alpha^\prime}\sum_{\mathbf{kk^\prime q}}W_{\alpha\alpha^{\prime}}a^+_{\alpha\mathbf{k}\uparrow}a^+_{\alpha-\mathbf{k+q}\downarrow}a_{\alpha^\prime-\mathbf{k^\prime+q}\downarrow}a_{\alpha^\prime\mathbf{k^\prime}\uparrow},\qquad
\end{eqnarray}
where $\tilde{\epsilon}_\alpha=\epsilon_\alpha-\mu$ is the electron
energy in the band $\alpha=1,2$ relative to the chemical potential $\mu$;
$V$ is the volume of superconductor and
$W_{\alpha\alpha^{\prime}}$ is the
matrix elements of intraband ($\alpha=\alpha^\prime$) or interband
($\alpha\neq\alpha^\prime$) pair transfer interaction. It is supposed that the chemical potential is located in the region of the bands
overlapping. We assume that (effective) electron-electron
interactions are nonzero only in the layer $\mu\pm\hbar\omega_\mathrm{D}$ and $W_{\alpha\alpha^{\prime}}$ is independent on electron wave
vector in this layer. For simplicity we take $W_{12}=W_{21}$.

We calculate the partition function $Z=\mathrm{Sp}\exp\left(\frac{-H}{k_\mathrm{B}T}\right)$ for the macroscopic system by means of
Hubbard-Stratonovich transformation \cite{Hubbard,*Stratonovich}. For  $W^2=W_{11}W_{22}-W_{12}^2>0$ and for real order parameters $\delta_\alpha$ the static path approximation reads as
\begin{eqnarray}\label{e2}
&&Z=\int\limits_{-\infty}^{\infty}e^{-\frac{\tilde{F}}{k_\mathrm{B}T}}\ud\delta_{1\mathbf{0}}\ud\delta_{2\mathbf{0}}\prod_{\mathbf{k>0}}\ud\delta^\prime_{1\mathbf{k}}\ud\delta^{\prime\prime}_{1\mathbf{k}}\ud\delta^\prime_{2\mathbf{k}}\ud\delta^{\prime\prime}_{2\mathbf{k}},\\
&&\tilde{F}=\tilde{F}_\mathrm{n}\nonumber\\
&&+\sum_{\alpha=1}^2\int
\left( a_\alpha\delta_\alpha^2+\frac{b_\alpha}{2}\delta_\alpha^4+K_\alpha(\nabla\delta_\alpha)^2-c\delta_\alpha\delta_{3-\alpha}\right)\ud V.\label{e3}\qquad
\end{eqnarray}
Here integration variables are treated as real and imaginary parts of Fourier components for non-equilibrium order parameters $\delta_\alpha(\mathbf{r})=\sum_\mathbf{k}\delta_{\alpha\mathbf{k}}e^{i\mathbf{kr}}$, $\tilde{F}$ is non-equilibrium free energy of inhomogeneous system and $\tilde{F}=\tilde{F}_\mathrm{n}$ in the absence of superconductivity. We do not expand the coefficients 
\begin{equation}\label{e4}
a_\alpha=\frac{W_{3-\alpha,3-\alpha}}{W^2}-\rho_\alpha\ln\frac{1.13\hbar\omega_\mathrm{D}}{k_\mathrm{B}T},
\end{equation}
$b_\alpha=\frac{0.11\rho_\alpha}{(k_\mathrm{B}T)^2}$, $c=\frac{W_{12}}{W^2}$ and $K_\alpha=\frac{0.02\rho_\alpha\hbar^2v_{F\alpha}^2}{(k_\mathrm{B}T)^2}$ in powers of $T-T_\mathrm{c}$, which allows us to apply free energy in the form (\ref{e3}) substantially farther from critical temperature $T_\mathrm{c}$. Note that the coefficient $K_\alpha$ used alludes isotropic situation.

The homogeneous equilibrium state is defined by the minimization $\frac{\delta\tilde{F}}{\delta\delta_\alpha}\big|_{\delta_\alpha=\Delta_\alpha}=0$, which gives us the set of equations for coupled homogeneous order parameters $\Delta_{1,2}$, namely
\begin{equation}\label{e5}
a_\alpha\Delta_\alpha+b_\alpha\Delta_\alpha^3=c\Delta_{3-\alpha},
\end{equation}
One should also take into account the relation between phases in equilibrium $\mathrm{sqn}(c\Delta_1\Delta_2)=+1$. The critical point is determined by condition $a_1(T_\mathrm{c})a_2(T_\mathrm{c})=c^2$, which has two solutions $T_{\mathrm{c}\pm}$ and $T_{\mathrm{c}-}>T_{\mathrm{c}+}$. If $T_{\mathrm{c}\alpha}$ are intrinsic transition temperatures in the bands and $T_{\mathrm{c}1}>T_{\mathrm{c}2}$, then for $W_{12}\to0$ we have $T_{\mathrm{c}-}\to T_{\mathrm{c}1}$ and $T_{\mathrm{c}+}\to T_{\mathrm{c}2}$. Note that in the system with coupled bands there is only one phase transition point $T_\mathrm{c}=T_{\mathrm{c}-}$.

Now we linearise functional $\tilde{F}$ near homogeneous state with free energy $F_\mathrm{h}$ by assuming 
$\delta_\alpha(\mathbf{r})=\Delta_\alpha+\eta_\alpha(\mathbf{r})$. By means of complex Fourier components $\eta_{\alpha\mathbf{k}}$ we have
\begin{eqnarray}\label{e6}
&&\tilde{F}=F_\mathrm{h}+V\sum_{\alpha=1}^2\bigg( A_{\alpha\mathbf{0}}\eta_{\alpha\mathbf{0}}^2-c\eta_{\alpha\mathbf{0}}\eta_{3-\alpha\mathbf{0}}+\nonumber\\
&&2\sum_\mathbf{k>0}\Big( A_{\alpha\mathbf{k}}\big(\eta_{\alpha\mathbf{k}}^{\prime
2}+\eta_{\alpha\mathbf{k}}^{\prime\prime
2}\big)-c\big(\eta_{\alpha\mathbf{k}}^\prime\eta_{3-\alpha\mathbf{k}}^\prime+\eta_{\alpha\mathbf{k}}^{\prime\prime}\eta_{3-\alpha\mathbf{k}}^{\prime\prime}\big)\Big)\bigg),\qquad
\end{eqnarray}
where $A_{\alpha\mathbf{k}}=A_\alpha+K_\alpha\mathbf{k}^2$ and $A_\alpha=a_\alpha+3b_\alpha\Delta_\alpha^2\geq0$. Note that due to interband pairing there appear non-diagonal terms in the quadratic form (\ref{e6}). Statistics for the equilibrium state fluctuations is determined by the distribution function $e^{-\frac{\tilde{F}}{k_\mathrm{B}T}}$ normalized to $Z$. By using Gaussian approximation (\ref{e6}) we calculate mean values $\langle\eta_{\alpha\mathbf{k}}\eta_{\alpha^\prime\mathbf{k}}^\ast
\rangle$ and then correlation functions $\Gamma_{\alpha\alpha^\prime}(\mathbf{r-r^\prime})=\sum_{\mathbf{k}}\langle\eta_{\alpha\mathbf{k}}\eta_{\alpha^\prime\mathbf{k}}^\ast
\rangle
e^{i\mathbf{k(r-r^\prime)}}$ for the order parameter fluctuations considered at different points separated by distance $|\mathbf{r-r^\prime}|\neq0$. We obtain $\Gamma_{\alpha\alpha^\prime}=\Gamma_{\alpha\alpha^\prime}^++\Gamma_{\alpha\alpha^\prime}^-$, where
\begin{equation}\label{e7}
\Gamma_{\alpha\alpha}^\pm=\mp\frac{k_\mathrm{B}T}{8\pi K_\alpha}\frac{\xi_\mp^2(\xi_\pm^2-\xi_{3-\alpha}^2)}{\xi_{3-\alpha}^2(\xi_-^2-\xi_+^2)}\frac{\exp\left(-\frac{|\mathbf{r-r^\prime}|}{\xi_\pm}\right)}{|\mathbf{r-r^\prime}|},
\end{equation}
and
\begin{equation}\label{e8}
\Gamma_{12}^\pm=\mp\frac{k_\mathrm{B}T}{8\pi K_1K_2}\frac{\xi_+^2\xi_-^2c}{\xi_-^2-\xi_+^2}\frac{\exp\left(-\frac{|\mathbf{r-r^\prime}|}{\xi_\pm}\right)}{|\mathbf{r-r^\prime}|}.
\end{equation}
Note also that $\Gamma_{12}=\Gamma_{21}$. In the Eqs. (\ref{e7})-(\ref{e8}) we have introduced $\xi_\alpha^2=\frac{K_\alpha}{A_\alpha}$ and the correlation lengths $\xi_\pm$ are given by
\begin{equation}\label{e9}
\xi_\pm^2=\frac{2\xi_1^2\xi_2^2}{\xi_1^2+\xi_2^2\pm\sqrt{(\xi_1^2-\xi_2^2)^2+4\xi_1^2\xi_2^2\frac{c^2}{A_1A_2}}}.
\end{equation}
These quantities have the following properties. For finite interband pairing $\xi_->\xi_+>0$. In the temperature region where $\xi_1>\xi_2$ one has $\xi_->\xi_1$ and $\xi_+<\xi_2$. For opposite case $\xi_1<\xi_2$ we get $\xi_->\xi_2$ and $\xi_+<\xi_1$. As a result, $\Gamma_{\alpha\alpha}^\pm>0$. However, $\mathrm{sqn}(\Gamma_{12}^\pm)=\mp\mathrm{sqn}(c)$, i.e. depending on the sign of interband constant one contribution in $\Gamma_{12}$ becomes negative.

The characteristics $\xi_\pm$ define the size of the region, where the order parameter fluctuations are significantly correlated. In fact, these length scales appear in the exponents despite the band index taken for the correlation functions, i.e. $\xi_\pm$ describe joint superconducting state rather than individual bands. We note also that $\xi_\pm$ coincide with the correlation lengths \cite{ord1} found by means of inhomogeneous gap equations.
\begin{figure*}
\resizebox{1.7\columnwidth}{!}{\includegraphics{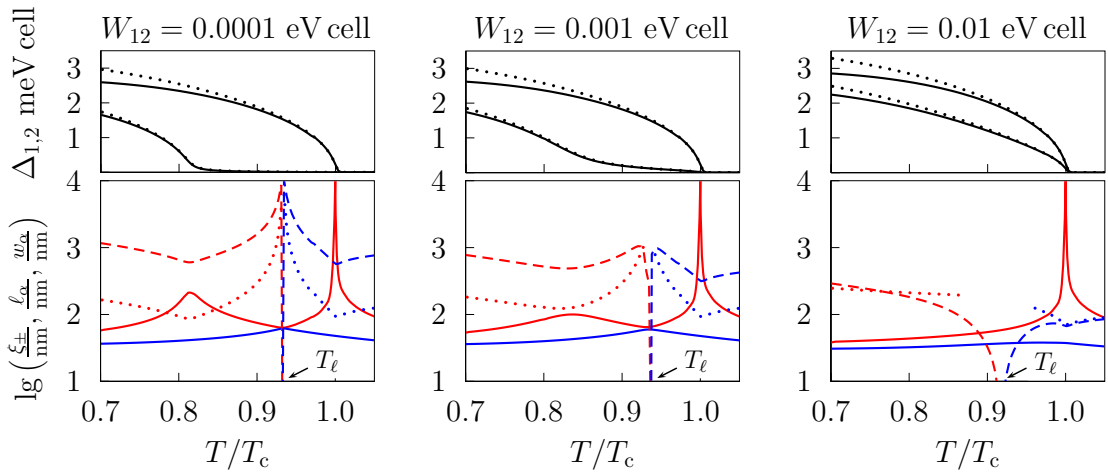}}
\caption{Above: The plots of the gaps $\Delta_\alpha$ as a solution of Eqs. (\ref{e5}) (solid) and derived microscopically (dotted) \textit{vs} temperature for various interband couplings $W_{12}$. Below: The $\log$ plots of $\xi_-$ (solid red), $\xi_+$ (solid blue), $\ell_1$ (dashed red), $\ell_2$ (dashed blue), $w_1(70\%)$ (dotted red) and $w_2(70\%)$ (dotted blue)  \textit{vs} temperature for same $W_{12}$.}\label{f1}
\end{figure*}
\section{Results and discussions}
\subsection{Correlation lengths}
The presence of interacting order parameters makes the coherence properties of the two-band system quite different from the corresponding characteristics in single-band superconductors. To analyse the physics of one-band case one should take 
$c\to0$. In this limit $\xi_\pm\to\xi_\alpha|_{c=0}$, i.e. one obtains two separate correlation length attributed to the band $\alpha=1,2$. Each length diverges at its own point given by intrinsic transition temperature $T_{\mathrm{c}\alpha}$. Note that $\xi_-\to\xi_1|_{c=0}$ and $\xi_+\to\xi_2|_{c=0}$ in the temperature region where $\xi_1|_{c=0}>\xi_2|_{c=0}$, however, $\xi_-\to\xi_2|_{c=0}$ and $\xi_+\to\xi_1|_{c=0}$ for the temperatures where $\xi_1|_{c=0}<\xi_2|_{c=0}$. Further we assume for specificity  $T_{\mathrm{c}2}<T_{\mathrm{c}1}$, i.e the condition $\xi_1|_{c=0}<\xi_2|_{c=0}$ corresponds to the lower temperature region, while $\xi_1|_{c=0}>\xi_2|_{c=0}$ to the higher temperatures in the superconducting state.

Non-zero coupling between bands modifies drastically the trivial physics of two non-interacting condensates. The coherency is described by lengths $\xi_\pm$ which become tricky combinations of band characteristics $\xi_\alpha$, see Eq. (\ref{e9}). To illustrate the evolution of $\xi_\pm$ with model parameters we fix intraband ones: $W_{11,22}=0.3\mathrm{\  eV\,cell}$, $\rho_{1,2}=(1,0.94)\mathrm{\ (eV\,cell)}^{-1}$, $v_{\mathrm{F}1,2}=(5,5.104)\cdot10^5\mathrm{\ m/s}$, $\textnormal{cell}=0.1\mathrm{\ nm}^3$. For these values $T_{\mathrm{c}2}=0.81T_{\mathrm{c}1}$. We also assume parabolic electron spectrum where $\frac{\rho_2}{\rho_1}=\big(\frac{v_{\mathrm{F}1}}{v_{\mathrm{F}2}}\big)^3$.

Fig. \ref{f1} shows temperature dependencies for correlation lengths together with the evolution of homogeneous gaps calculated numerically as interband coupling increases. We see that $\xi_-$ and $\xi_+$ as functions of the temperature are remarkably different. First, the length $\xi_-$ behaves critically diverging at phase transition point $T_\mathrm{c}$. At the same time $\xi_+$ remains finite. Second, $\xi_-$ can change below $T_\mathrm{c}$ very non-monotonically, while the temperature dependence of $\xi_+$ is substantially weaker \cite{babaev2,ord1}. The appearance of additional maximum in superconducting phase for $\xi_-$ is strongly supported by the smaller values of $W_{12}$, representing the memory effect about criticality in the band $\alpha=2$. The position of this maximum is correlated with the inflection point of the smaller gap which takes place in the vicinity of $T_{\mathrm{c}2}$. As was pointed earlier \cite{babaev1}, the non-monotonicity of the critical coherence length elucidates the temperature behaviour of the gaps healing length \cite{shanenko} and vortex size \cite{shanenko4} in a superconductor with weakly interacting bands.

One can argue that the scheme based on the expansion (\ref{e3}) is applicable only close to critical point. We note that the  coefficients (\ref{e4}) taken allow us to go essentially farther below $T_\mathrm{c}$. For the comparison we have plotted in Fig. \ref{f1} homogeneous gaps calculated numerically by means of microscopic theory. The latter are approximated by the solutions of system (\ref{e5}) very well in the temperature region considered.

Due to the definition of the critical point $a_1(T_\mathrm{c})a_2(T_\mathrm{c})=c^2$ and the relation $A_\alpha(T_\mathrm{c})=a_\alpha(T_\mathrm{c})$ one obtains
\begin{equation}\label{e10}
\xi_+^2(T_\mathrm{c})=\frac{\xi_1^2(T_\mathrm{c})\xi_2^2(T_\mathrm{c})}{\xi_1^2(T_\mathrm{c})+\xi_2^2(T_\mathrm{c})},
\end{equation}
and zero for the denominator of $\xi_-(T_\mathrm{c})$, i.e. the latter length diverges precisely at $T_\mathrm{c}$. This implies that only length scale $\xi_-$ can be attributed directly to the superconducting phase transition in a two-band model. In the vicinity of $T_\mathrm{c}$ we get
\begin{equation}\label{e11}
\xi_-^2=\Bigg\{\begin{array}{ccc}c^2\frac{\xi_1^2(T_\mathrm{c})+\xi_2^2(T_\mathrm{c})}{\rho_1a_2(T_\mathrm{c})+\rho_2a_1(T_\mathrm{c})}\frac{T_\mathrm{c}}{T-T_\mathrm{c}},&T>T_\mathrm{c}\\
\frac{c^2}{2}\frac{\xi_1^2(T_\mathrm{c})+\xi_2^2(T_\mathrm{c})}{\rho_1a_2(T_\mathrm{c})+\rho_2a_1(T_\mathrm{c})}\frac{T_\mathrm{c}}{T_\mathrm{c}-T},&
T<T_\mathrm{c}\end{array}.
\end{equation}
The expressions (\ref{e10})-(\ref{e11}) one meets also in the literature \cite{poluektov,kogan}.

\begin{figure}
\resizebox{1.0\columnwidth}{!}{\includegraphics{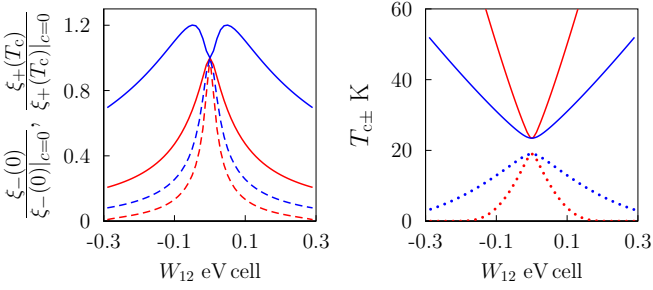}}
\vspace{-0.5cm}\caption{Left: The plots of $\xi_-(0)$ (solid) and $\xi_+(T_\mathrm{c})$ (dashed) normalized to their values at $W_{12}=0$ \textit{vs} interband coupling $W_{12}$. Right: The plots of $T_\mathrm{c}=T_{\mathrm{c}-}$ (solid) and $T_{\mathrm{c}+}$ (dotted) \textit{vs} $W_{12}$. Red curves correspond to the parameters given in text, blue ones to the modified parameters in band $\alpha=2$, namely, $W_{22}=2.26\mathrm{\  eV\,cell}$, $\rho_2=0.13\mathrm{\ (eV\,cell)}^{-1}$, $v_{\mathrm{F}2}=10^6\mathrm{\ m/s}$. For these values we have the same ratio $T_{\mathrm{c}2}=0.81T_{\mathrm{c}1}$. }\label{f2}
\end{figure}

Next we denote the factor in Eq. (\ref{e11}) by $\xi_-^2(0)$, the value of the formula (\ref{e11}) at $T=0$, and analyse $\xi_-(0)$ and $\xi_+(T_\mathrm{c})$ as the functions of interband interaction. Fig. \ref{f2} shows these dependencies for different sets of intraband parameters. Analytic consideration indicates that $\xi_+(T_\mathrm{c})$ always decreases with $|W_{12}|$, while $\xi_-(0)$ can pass through a maximum at some finite value of $W_{12}$. We interpret this feature as follows. The one-band limit $c=0$ gives $T_\mathrm{c}=T_{\mathrm{c}1}$ and $a_1(T_{\mathrm{c}1})=0$. As a result, $\xi_-^2|_{c=0}=\frac{K_1(T_{\mathrm{c}1})}{2\rho_1}\frac{T_{\mathrm{c}1}}{T_{\mathrm{c}1}-T}$ for $T<T_{\mathrm{c}1}$. This is standard one-band expression for the squared correlation length expanded near critical point. The factor $\xi_-^2(0)|_{c=0}=\frac{K_1(T_{\mathrm{c}1})}{2\rho_1}$ is proportional to $\frac{1}{T_{\mathrm{c}1}^2}$, i.e. $\xi_-(0)|_{c=0}
$ decreases with the critical temperature increase and \textit{vice versa}. In two-band system $T_\mathrm{c}$ always grows with $W_{12}$ (see Fig. \ref{f2}) and one can expect the reduction of $\xi_-(0)$ with increase of $|W_{12}|$ by analogy with one-component case. However, in two-component situation, especially for weak interband couplings, the memory effect related to the lower intrinsic phase transition is strong. The latter is characterized by the temperature $T_{\mathrm{c}+}$ which always decreases with $|W_{12}|$ (see Fig. \ref{f2}). By analogy with one-band case it can lead to the rise of $\xi_-(0)$. Thus, there are two opposite tendencies associated with the temperatures $T_{\mathrm{c}\pm}$ which govern the behaviour of $\xi_-(0)$ as a function of interband coupling. By analysing this competition analytically we find that, if 
\begin{equation}\label{e12}
\frac{v_{\mathrm{F}2}}{v_{\mathrm{F}1}}<\sqrt{1+2\frac{\rho_1W_{11}-\rho_2W_{22}}{(\rho_1W_{11})^2}},
\end{equation}
$\xi_-(0)$ has maximum, whereas for opposite sign in Eq. (\ref{e12}) the function $\xi_-(0)$ has minimum at $W_{12}=0$. We believe that non-monotonicity of $\xi_-(0)$ is clear footprint of the two-band nature near critical point.

One comment should be made about non-critical coherence length . The quantity $\xi_+^2$ is always finite and decreases as the strength of interband interaction increases, crossing zero at $W^2=0$. At the same time, there is natural lower bound for coherence lengths in Ginzburg-Landau theory defined by the microscopic length scales (Cooper pair size in the bands) $\frac{\hbar v_{\mathrm{F}\alpha}}{k_\mathrm{B}T_\mathrm{c}\pi}$. The latter guarantees the smallness of the gradient term in Ginzburg-Landau expansion. To estimate the maximal value of $\xi_+$ we use Eq. (\ref{e10}) for $c=0$. We find
\begin{equation}\label{e13}
\xi_+(T_\mathrm{c})|_{c=0}=\xi_2(T_{\mathrm{c}1})|_{c=0}\sim\frac{1}{\sqrt{\rho_1W_{11}-\rho_2W_{22}}}.
\end{equation}
Consequently, the value of $\xi_+(T_\mathrm{c})$ can be magnified when $T_{\mathrm{c}2}$ approaches $T_{\mathrm{c}1}$. In this process non-critical coherence length can surpass microscopic lengths \cite{ord1}, i.e. two length scales of coherency found are meaningful even in the standard two-band Ginzburg-Landau model for relevant parameters. To overcome the restriction related to the microscopic lengths one should take into account the higher terms of the gradient expansion in the Ginzburg-Landau approach. In this way one gets better agreement with microscopic theory \cite{babaev1}. However, the theory based on two-band Eilenberger equations also predicts the disappearance of non-critical length for strong interband pairings at $W^2\approx0$ \cite{babaev2}. The absence of the real non-critical correlation length may signal spatial periodicity of fluctuations of two-gap superconductivity \cite{ord2}.

\subsection{Correlation functions}
Interaction between bands results more complicated structure of correlation functions as compared to the case of decoupled bands for which
\begin{equation}\label{e14}
\Gamma_{\alpha\alpha}=\frac{k_\mathrm{B}T}{8\pi K_\alpha|\mathbf{r-r^\prime}|}e^{-\frac{|\mathbf{r-r^\prime}|}{\xi_\alpha|_{c=0}}},\qquad\qquad\Gamma_{12}=0.
\end{equation}
Next we discuss the correlation functions for non-vanishing interband pairings.

First, we consider different spatial regions. For shorter distances $|\mathbf{r-r^\prime}|\ll\xi_+<\xi_-$ (denote as "sd") we obtain from Eqs. (\ref{e7})-(\ref{e8})
\begin{eqnarray}\label{e15}
&&\Gamma_{\alpha\alpha}^\mathrm{sd}\approx\frac{k_\mathrm{B}T}{8\pi K_\alpha|\mathbf{r-r^\prime}|},\qquad\Gamma_{12}^\mathrm{sd}\approx\frac{k_\mathrm{B}Tc}{8\pi K_1K_2}\frac{\xi_+\xi_-}{\xi_++\xi_-}.\qquad
\end{eqnarray}
In this fully correlated case the functions $\Gamma_{\alpha\alpha^\prime}$ are maximal. If $\xi_1(T_\mathrm{c})\gg\xi_2(T_\mathrm{c})$, then near $T_\mathrm{c}$ the main contribution to $\Gamma_{11}^\mathrm{sd}$ stems from critical, while to $\Gamma_{22}^\mathrm{sd}$ from non-critical channel of coherency and \textit{vice versa}. Note that condition $\xi_1(T_\mathrm{c})\gg\xi_2(T_\mathrm{c})$ is supported by the smaller interband interaction.

\begin{figure}
\resizebox{1.0\columnwidth}{!}{\includegraphics{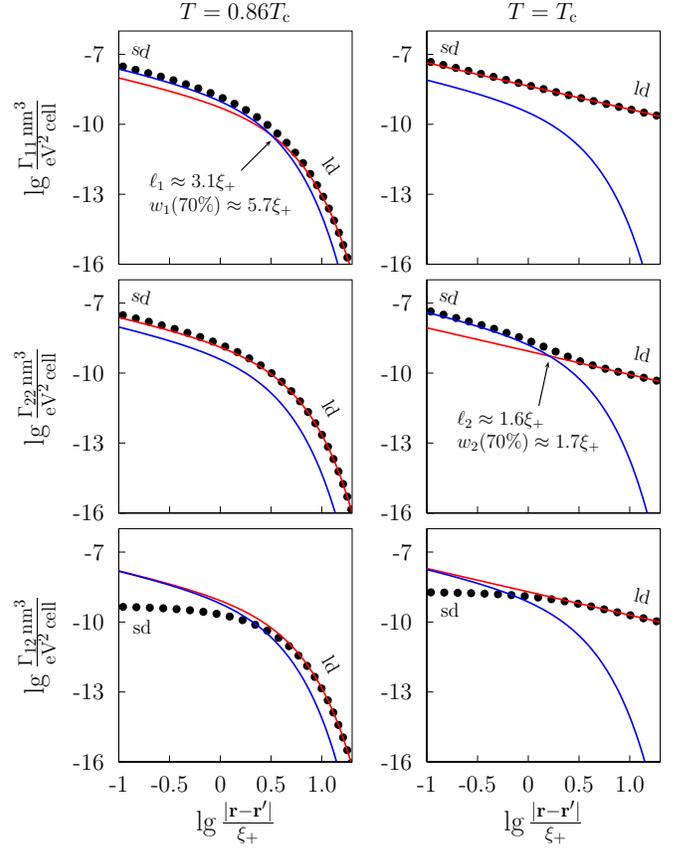}}
\vspace{-0.5cm}\caption{The log plots of $\Gamma_{\alpha\alpha^\prime}$ (dots) together with corresponding contributions $\Gamma_{\alpha\alpha^\prime}^-$ (red), $\Gamma_{\alpha\alpha}^+$ and $|\Gamma_{12}^+|$ (blue) \textit{vs} distance $\mathbf{r-r^\prime}$ for $T=0.86T_\mathrm{c}$ (left column) and  $T=T_\mathrm{c}$ (right column). Here $W_{12}=0.01\mathrm{\ eV\,cell}$ and intraband parameters as in text. At $T_\mathrm{c}$ we have $\frac{\xi_1}{\xi_2}\approx2.2$. The regimes "sd" and "ld" are discussed in text. }\label{f3}
\end{figure}
For larger distances $\xi_+<\xi_-\ll|\mathbf{r-r^\prime}|$ (denote as "ld") the functions $\Gamma_{\alpha\alpha^\prime}^\mathrm{ld}$ are defined mostly by critical contributions and they vanish. By approaching $T_\mathrm{c}$ we have in this regime $\xi_+\ll|\mathbf{r-r^\prime}|\ll\xi_-$ and
\begin{eqnarray}\label{e16}
&&\Gamma_{\alpha\alpha}^\mathrm{ld}\approx\frac{k_\mathrm{B}T_\mathrm{c}}{8\pi\mathcal{K}_\alpha(T_\mathrm{c})|\mathbf{r-r^\prime}|},\qquad\mathcal{K}_\alpha=K_\alpha\frac{\xi_1^2+\xi_2^2}{\xi_\alpha^2},\qquad\\
&&\Gamma_{12}^\mathrm{ld}\approx\frac{k_\mathrm{B}T_\mathrm{c}c}{8\pi K_1(T_\mathrm{c})K_2(T_\mathrm{c})}\frac{\xi_+^2(T_\mathrm{c})}{|\mathbf{r-r^\prime}|}.\label{e17}
\end{eqnarray}
Thus, at critical point $\Gamma_{12}$ changes in space from constant value $\Gamma_{12}^\mathrm{sd}$ to the function $\Gamma_{12}^\mathrm{ld}$ which decreases linearly with $\log\frac{|\mathbf{r-r^\prime}|}{\xi_+}$. The disagreement between $\mathcal{K}_\alpha$ and $K_\alpha$ characterises the behaviour of $\Gamma_{\alpha\alpha}$. If $\xi_1(T_\mathrm{c})\gg\xi_2(T_\mathrm{c})$, then $\Gamma_{11}^\mathrm{sd}$ and $\Gamma_{11}^\mathrm{ld}$ are the very same function at $T_\mathrm{c}$, but there is remarkable difference in the dependencies $\Gamma_{22}^\mathrm{sd}$ and $\Gamma_{22}^\mathrm{ld}$ related to the change of the dominant coherency channel from non-critical to critical one. This transformation is also noticeable in Fig. \ref{f3}, and it is supported by the weaker interband couplings. For opposite situation $\xi_1(T_\mathrm{c})\ll\xi_2(T_\mathrm{c})$ we have different dependencies for $\Gamma_{11}^\mathrm{sd}$ and $\Gamma_{11}^\mathrm{ld}$, but same for  $\Gamma_{22}^\mathrm{sd} $ and $\Gamma_{22}^\mathrm{ld}$. Therefore, the changes in spatial functionality of the correlation functions are intrinsic for two-gap superconductors.

To estimate the efficiency of different correlation channels we find the crossover points where $\Gamma_{\alpha\alpha}^+=\Gamma_{\alpha\alpha}^-$ and crossover width.  For $\Gamma_{11}$ the crossover takes place at distance
\begin{equation}\label{e18}
\ell_1=\frac{\xi_+\xi_-}{\xi_--\xi_+}\ln\frac{1-\frac{\xi_2^2}{\xi_+^2}}{\frac{\xi_2^2}{\xi_-^2}-1}=-\ell_2,
\end{equation}
where $\ell_2$ is the corresponding parameter derived for $\Gamma_{22}$. For fixed temperature the crossover takes place only in the behaviour of certain correlation function, $\Gamma_{11}$ or $\Gamma_{22}$.

From definition (\ref{e18}) follows that there is the temperature $T_\ell$ defined by the condition $\xi_1(T_\ell)=\xi_2(T_{\ell})$ where $\ell_\alpha$ goes to zero. The position of $T_\ell$ is sensitive to the model parameters. If $v_{\mathrm{F}1}>v_{\mathrm{F}2}$ one has $T_\ell<T_\mathrm{c}$. However, for $v_{\mathrm{F}1}<v_{\mathrm{F}2}$ there is a value
\begin{equation}\label{e19}
|W_{12}|=\frac{v_{\mathrm{F}1}v_{\mathrm{F}2}}{v_{\mathrm{F}2}^2-v_{\mathrm{F}1}^2}\frac{\rho_1W_{11}-\rho_2W_{22}}{\sqrt{\rho_1\rho_2}},
\end{equation}
for which $T_\ell=T_\mathrm{c}$, and for stronger interband interaction $T_\ell>T_\mathrm{c}$. Next, for $\xi_1(T)<\xi_2(T)$ we have $\ell_1>0$ and \textit{vice versa}. At $T_\mathrm{c}$ one obtains $\ell_2=\xi_+\ln\frac{\xi_1^2}{\xi_2^2}$. Fig. \ref{f1} shows also that $\ell_\alpha$ can substantially exceed $\xi_-$, especially for nearly decoupled bands. In this case the value $\Gamma_{\alpha\alpha}(\ell_\alpha)$ is vanishing, i.e. non-critical channel dominates for all reasonable distances.

Finally, the relative contribution $\frac{\Gamma_{\alpha\alpha}^+}{\Gamma_{\alpha\alpha}}$ decreases and $\frac{\Gamma_{\alpha\alpha}^-}{\Gamma_{\alpha\alpha}}$ increases with distance. If $\ell_\alpha\geq0$ these functions cross at $\ell_\alpha$, i.e. $\frac{\Gamma_{\alpha\alpha}^+}{\Gamma_{\alpha\alpha}}\geq\frac{\Gamma_{\alpha\alpha}^-}{\Gamma_{\alpha\alpha}}$ for tiny $|\mathbf{r-r^\prime}|$. We define the width $w_\alpha$ of crossover region as the size of the spatial area  around $\ell_\alpha$ where $\frac{\Gamma_{\alpha\alpha}^+}{\Gamma_{\alpha\alpha}}$ and $\frac{\Gamma_{\alpha\alpha}^-}{\Gamma_{\alpha\alpha}}$ simultaneously do not exceed  fixed percentage $p>50\%$. We find
\begin{equation}\label{e20}
w_\alpha(p)=\frac{\xi_+\xi_-}{\xi_--\xi_+}\ln\frac{p^2}{(1-p)^2}.
\end{equation}
At $T_\mathrm{c}$ one obtains $w_\alpha\sim\xi_+$, i.e. the crossover between contributions in $\Gamma_{\alpha\alpha}$ takes place on the distance defined by non-critical length scale. The width $w_\alpha$ shrinks at $T_\mathrm{c}$ with increase of interband coupling. For nearly decoupled bands $w_1\sim\xi_+$ holds also in the vicinity of $T_{\mathrm{c}2}$ (see Fig. \ref{f1}), however, $w_1$ grows with $W_{12}$ at that temperatures. In Fig. \ref{f1} we distinguish clear crossover from close coexistence of contributions up to $\ell_\alpha$. The latter occurs in the region around $T_\ell$ which widens as interband interaction increases. 

\section{Conclusions} The evolution of correlation functions for two-band superconductivity indicates the presence of two distinct channels of coherency described by the critical (divergent at critical point) and non-critical (finite at critical point) correlation lengths. Although these characteristics are not related directly to the bands, two-component nature manifests itself e.g. in the non-monotonicities of critical length scale as a function of the temperature and the strength of interband interaction. The features of the competition between coherency channels involved depend on the temperature as well as model parameters, e.g. coupling between bands or Fermi velocities.

\section{Acknowledgement} This study was supported by the European Union through the European Regional Development Fund (Centre of Excellence "Mesosystems: Theory and Applications", TK114) and by the Estonian Science Foundation (Grant No 8991).

\bibliography{2bandLength2}
\end{document}